\documentclass[twocolumn,superscriptaddress, floatfix, showpacs, aps, prb, 10pt]{revtex4-2}
\usepackage{hyperref}
\usepackage{physics}
\usepackage{color}
\usepackage{graphicx}
\usepackage{dcolumn} 
\usepackage{bm}

\begin{document}

\makeatletter
\newcommand{\globalcolor}[1]{%
  \color{#1}\global\let\default@color\current@color
}
\makeatother

\newcommand{\nm}{\mathbin{\phantom{-}}}

\title{Fermi surface and topology of multiband superconductor BeAu}

\author{Riccardo Vocaturo}
\affiliation{Leibniz Institute for Solid State and Materials Research (IFW) Dresden, Helmholtzstrasse 20, 01069 Dresden, Germany}

\author{Klaus Koepernik}
\affiliation{Leibniz Institute for Solid State and Materials Research (IFW) Dresden, Helmholtzstrasse 20, 01069 Dresden, Germany}

\author{D\'aniel Varjas}
\affiliation{Leibniz Institute for Solid State and Materials Research (IFW) Dresden, Helmholtzstrasse 20, 01069 Dresden, Germany}
\affiliation{W{\"u}rzburg-Dresden Cluster of Excellence ct.qmat, Helmholtzstrasse 20, 01069 Dresden, Germany}
\affiliation{Department of Theoretical Physics, Institute of Physics,
Budapest University of Technology and Economics, M\H{u}egyetem rkp. 3., 1111 Budapest, Hungary}

\author{Oleg Janson}
\affiliation{Leibniz Institute for Solid State and Materials Research (IFW) Dresden, Helmholtzstrasse 20, 01069 Dresden, Germany}

\author{Maia G. Vergniory}
\affiliation{Regroupement Qu\'{e}b\'{e}cois sur les Mat\'{e}riaux de Pointe (RQMP), Quebec H3T 3J7, Canada}
\affiliation{Donostia International Physics Center, 20018 Donostia-San Sebastian, Spain}
\affiliation{Département de Physique et Institut Quantique,
Université de Sherbrooke, Sherbrooke, J1K 2R1, Québec, Canada.}

\author{Jeroen van den Brink}
\affiliation{Leibniz Institute for Solid State and Materials Research (IFW) Dresden, Helmholtzstrasse 20, 01069 Dresden, Germany}
\affiliation{W{\"u}rzburg-Dresden Cluster of Excellence ct.qmat, Helmholtzstrasse 20, 01069 Dresden, Germany}
\affiliation{Institute of Theoretical Physics, Technische Universit\"{a}t Dresden, 01062 Dresden, Germany}

\date{\today}

\begin{abstract}
	The chiral material BeAu  was recently identified as a multiband
	type-I superconductor with a critical temperature of 3.2~K.  As a member of the B20
	crystal family (space group $P2_13$), its band structure hosts multifold
	fermions at high-symmetry points, unpaired Weyl points
	and even nodal surfaces. This renders BeAu an appealing system to investigate
	the interplay between superconductivity and topology. 
	Here we present a comprehensive first-principles analysis of
	BeAu’s electronic structure focusing on its Fermi surface's topology and
	the implications for superconductivity. Together with the presence of
	four- and six-fold fermions at high-symmetry points, we identify several additional isolated Weyl points near the Fermi level. 
    We also determine the associated topological edge states---the surface Fermi arcs.
    Computing the Chern number associated to different Fermi
 surface sheets, we show that BeAu harbors a $\nu = 4$ topological superconducting phase in the presence of $s$-wave pairing of alternating sign ($s_\pm$ pairing).  Notably, we also identify
	a Fermi surface with a Chern number of +6; the highest value reported to date.
	Finally, our analysis reveals strong inhomogeneity in the orbital
	character of electronic states at the Fermi level, suggesting a link to the observed multigap
	superconductivity. 
\end{abstract}

\keywords{Weyl semimetals, multifold fermions, topological superconductivity}
\maketitle

\section{Introduction}

\noindent 
The family of B20 compounds has been gaining significant attention in
recent years, presenting itself as a natural platform for investigating exotic
topological phases~\cite{Felser2024}.  Materials belonging to this family crystallize in the FeSi-type
lattice, with chiral space group $P2_{1}3$ (no. 198). This structure is
characterized by the absence of inversion as well as mirror symmetries, and the presence of 3 screw-rotation axes along the unit cell's main diagonals.
Such uncommon combination of crystalline symmetries provides ideal
conditions for realizing unconventional electronic features in the band
structure. For instance, symmetry-enforced multiband crossings occur at
high-symmetry points in the Brillouin zone (BZ), carrying topological charges as
large as five~\cite{Alpin2023}. These crossings, termed multifold
fermions~\cite{Barry2016,Robredo2024}, have been
extensively studied in recent years due to the possibility to yield extremely long surface Fermi arcs~\cite{Krieger2024},
second-harmonic generation~\cite{deJuan2017,Lu2022}, quantized circular
photogalvanic effect~\cite{Flicker2018}, and many more (non-linear) transport
properties~\cite{Ahn2023,Robredo2021,Orlova2023,Ni2020}. 

Yet, multifold fermions are not the only unconventional features in chiral crystals.
For example, space group 198 is among the few known cases where nodal surfaces
(NS) and isolated Weyl points (WPs) can be found. Here, these two elements appear
together and make it possible to circumvent the famous Nielsen–Ninomiya theorem,
allowing the existence of a single WP in a periodic lattice.
These types of excitations were first proposed in 2019~\cite{Wu2018,Yu2019} and experimentally observed two years later in another B20 compound,
PtGa~\cite{Ma2021}.

Another aspect of great interest lays in the appealing prospect of realizing
topological superconductivity. Qi \textit{et al.}~\cite{Qi2010} showed that
time-reversal-invariant Weyl semimetals can undergo a topological
superconducting transition if the pairing function changes sign between Fermi
surfaces (FSs) of different chirality. In the weak-coupling limit, the
topological invariant is given by:
\begin{equation} 
	\nu = \frac{1}{2} \sum_{i \in FS} C_i \, \text{sgn}(\Delta_i)
	\label{eq:invariant}
\end{equation}
where $\Delta_i$ is the superconducting gap on the $i$-th FS, and $C_i$ is its
Chern number, computed as the flux of Berry curvature through the $i$-th sheet.
In a conventional Weyl semimetal, $C_i = \pm 1$ for Fermi surfaces
enclosing a Weyl point and zero otherwise.
While a sign-changing order parameter is not typically expected in
conventional electron-phonon superconductors, Hosur \textit{et
al.}~\cite{Hosur2014} showed that second-order effects from Coulomb repulsion between
different FSs can stabilize such a state, even in the presence of purely local
interactions.
In this context, B20 compounds are interesting because they host multifold
fermions at high-symmetry points, giving rise to extended and well-separated
Fermi surfaces with large Chern numbers. Contrary, in
conventional Weyl semimetals, WPs are usually located close to one another, 
resulting in a smaller probability to observe a sign change of the superconducting gap across the corresponding pockets.
Significant attention has thus been devoted to the effect of
superconductivity on the multifold fermions at the $\Gamma$ and $R$
points~\cite{Mardanya2024,Huang2024,Lee2021}. In minimal toy models, it has been
predicted that $s_{\pm}$ pairing may even drive a topological
superconducting transition with invariant $\nu = 4$~\cite{Peccato2021}.

In this work we turn our attention to BeAu, which,
remarkably, exhibits a superconducting transition $T_c \approx
3.2$~K~\cite{Amon2018,Beare2019,Singh2019}, despite the poor superconducting
properties of both beryllium and gold. It is also one of only two known
superconductors in the B20 family (the other being
RhGe~\cite{TSVYASHCHENKO2016431}). Additionally, multiple experimental evidence
points toward the presence of multiband superconductivity this
system~\cite{Rustem2020,Rustem2020_2,Datta2022}.

The agreement between the experimentally observed $T_c$ and theoretical
calculations~\cite{arxiv} suggests that superconductivity in BeAu is primarily
driven by electron-phonon coupling from light Be atoms; a mechanism reminiscent
of high-pressure hydrides. However, the presence of two distinct
superconducting gaps leaves two important questions unanswered: which FSs are involved, and
what mechanism gives rise to their different gap magnitudes? Experimental data
further hints that one of the gaps is strongly coupled while the other is weakly
coupled~\cite{arxiv}, pointing towards a significant anisotropy in the interaction
strength. This raises the possibility that an $s_{\pm}$ pairing may emerge
despite the dominance of electron-phonon coupling, leading to a
topological superconducting phase.

Since previous theoretical studies~\cite{Peccato2021} were conducted at the
effective model level, inevitably neglecting material-specific aspects, we present here a comprehensive first-principles investigation of
the electronic structure of BeAu, with a particular focus on its Fermi surface.
After a general overview of the crystal (Sec.~\ref{sec:cry_str}) and electronic (Sec.~\ref{sec:e_str}) structure
we highlight the presence of known multifold fermions at the high-symmetry points (Sec.~\ref{sec:multifold})
and perform a systematic search throughout the BZ to identify WPs at
generic positions, finding multiple \textit{isolated} crossings in the proximity of
the Fermi level (Sec.~\ref{sec:unpaired_crossings}).  We also examine in depth the surface spectral properties,
discussing important aspects which have not been considered in a recent
study~\cite{Bohwmick2025}.
In Sec.~\ref{sec:fs} and Sec.~\ref{sec:sc}, we focus on the complex Fermi surface of this system, revealing
additional electron and hole pockets beyond those at $\Gamma$ and R points captured by
the effective model in Ref.~\cite{Peccato2021}.  Moreover, the presence of several isolated
crossings in the proximity of the Fermi level complicates the determination of
the topological charge of each sheet, which is not easily inferred from a 
counting argument alone. Consequently, to rigorously confirm the possibility of
a non-trivial topological invariant in the presence of $s_{\pm}$ superconductivity, we carry
out a careful numerical computation of Fermi-surface Chern numbers.

Our main result is a complete topological characterization of the Fermi surface
of BeAu, showing that it is consistent with the $\nu = 4$ topological
superconducting phase theoretically proposed in minimal
models~\cite{Peccato2021,Lee2021}. We also identify a Fermi surface sheet with a Chern
number of $C = 6$ which could, in principle, support even richer
superconducting phases. To the best of our knowledge, this is also the highest value reported for a
single Fermi surface. Finally, we find a highly anisotropic distribution of
atomic orbital character at the Fermi level, a factor that could influence the
electron-phonon interaction and help explain the origin of the observed multigap
behavior.

\section{Crystal structure}
\label{sec:cry_str}

\noindent The low-temperature phase of BeAu emerges below
860\,K~\cite{Singh2019}, crystallizing in the chiral FeSi-type
structure with the cubic unit cell parameter $a$ = 4.6699(4) \r{A} at room temperature~\cite{Amon2018}. This structure type features sevenfold coordination polyhedra and represents an approximation to icosahedral quasicrystals~\cite{Dmitrienko1994}.

The crystal structure belongs to the well-known space group \( P2_13 \)
(No.~198), which is chiral due to the absence of both inversion symmetry and
mirror planes.  Its symmetry operations include a \( C_3 \) rotational axis
along the [111] direction and three non-symmorphic screw rotations: \( \{2_1\left(100\right)
	\mid \frac{1}{2} \, \frac{1}{2} \, 0\} \), \( \{2_1\left(010\right) \mid 0 \,
\frac{1}{2} \, \frac{1}{2}\} \), and \( \{2_1\left(001\right) \mid \frac{1}{2} \, 0 \,
\frac{1}{2}\} \). In this work, we will refer to such screw rotations with the notation $\tilde C_i$ with $i =
x,\ y,\ z$.
As structural input for our calculations, we use the experimental room-temperature structure where Au and Be
occupy the $4a$ Wyckoff positions with coordinates \( x_{\text{Au}} = 0.8549 \)
and \( x_{\text{Be}} = 0.15 \)~\cite{Amon2018}.

\section{Electronic structure}
\label{sec:e_str}

\noindent To study the electronic structure of BeAu, we performed
full-relativistic density functional theory (DFT) calculations using the
full-potential local-orbital code FPLO~\cite{Koepernik1999} version 22.01-63.
For the exchange-correlation term, we employed the generalized gradient
approximation (GGA)~\cite{Perdew1997}. The BZ was sampled by a
$12\times12\times12$ mesh of $k$-points.  For slab calculations, a suitable
Wannier representation was obtained through the dedicated FPLO
module~\cite{Koepernik2023}. To do so, Kohn-Sham eigenstates were projected onto
a basis of 80 spinfull orbitals that comprises Be $2s$, $2p$ and Au $6s$, $5d$;
in an energy window between $-9$ and $5$~eV.

\begin{figure}[ht]
	\centering
	\includegraphics[width=0.48\textwidth]{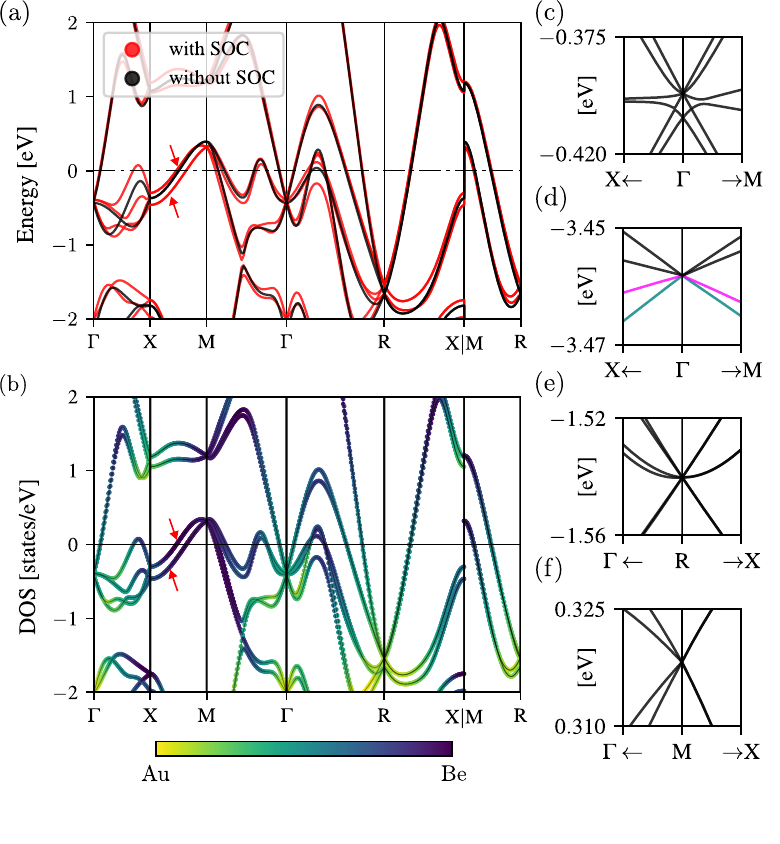}
	\caption{(a) Scalar- (without SOC) and full-relativistic (with SOC) GGA
	band structures of BeAu.  (b) Atomic-resolved full-relativistic band
	structure, highlighting the relative Au and Be contributions. (c-f)
	Zoom-in on the multifold band crossing at high-symmetry points carrying
	non-trivial Chern numbers (full-relativistic band structure). In (d) we color in pink the band carrying
	$C=5$ (see discussion in the main text for details).}
	\label{fig:e_str}
\end{figure}

The band structure with and without spin-orbit coupling (SOC) is shown in
Fig.~\ref{fig:e_str} (a) while in Fig.~\ref{fig:e_str} (b) we
resolve atomic contributions of each band at each momentum.
The density of states (DOS) have been previously published in
Ref.~\cite{Amon2018,arxiv}. In particular, in \cite{arxiv}, it was shown that Be
and Au atomic contribution are essentially equal around $E_F$. This is
consistent with Fig.~\ref{fig:e_str} (b), indicating strong hybridization of valence
electrons in this compound. However, the atomic-resolved band structure
evidences quantitative difference in the orbital character of some states. For
instance, the hole pocket around the M point has a remarkably higher Be contribution with respect to
the neighbouring bands crossing the Fermi level. 

The impact of SOC can also be understood comparing Fig.~\ref{fig:e_str}~(a) to
the orbital-resoved band structure in panel (b). As expected, Be, being a light element,
induces negligible effects, while much
wider band splitting are associated with regions of the BZ
with stronger Au contribution. For instance, comparing \(\Gamma\)--M and M--X
paths, we observe that along \(\Gamma\)--M, where Be
character dominates, the SOC-induced splitting is smaller. As we move toward
\(\Gamma\) along the \(\Gamma\)--X direction, the splitting increases, in line
with a higher Au contribution.
In Ref.~\cite{Bohwmick2025}, the weaker spin-orbit splitting of Be-dominated
states was invoked to explain the lower spin-Hall conductivity of this material
compared to the Weyl semimetal TaAs or the conventional metal Pt. However, the
same study also identified a significant amount of spin Berry
curvature near the Fermi level, which results in a much higher spin-Hall
conductivity relative to other B20 compounds.

\subsection{Topological multifold crossings and Fermi arcs}
\label{sec:multifold}

\noindent 
The band structure of BeAu features multiple examples of exotic crossings which
are not encountered in conventional Weyl semimetals and that are enforced by the symmetries of the crystal. 

While conventional WPs are described by a local Hamiltonian which
is adiabatically connected to $H_{\mathbf k} = \bm{\sigma} \cdot \mathbf k$, the
presence of additional crystalline symmetries allows the generalization to
higher dimensional intersections. Such objects will be described by an Hamiltonian of
the form $H_{\mathbf k} = \mathbf S \cdot \mathbf k$ with $\mathbf S =
(S_x,S_y,S_z)$ being some $n$-dimensional (iso)spin matrices.

\begin{figure*}[ht]
	\centering
	\includegraphics[width=1\textwidth]{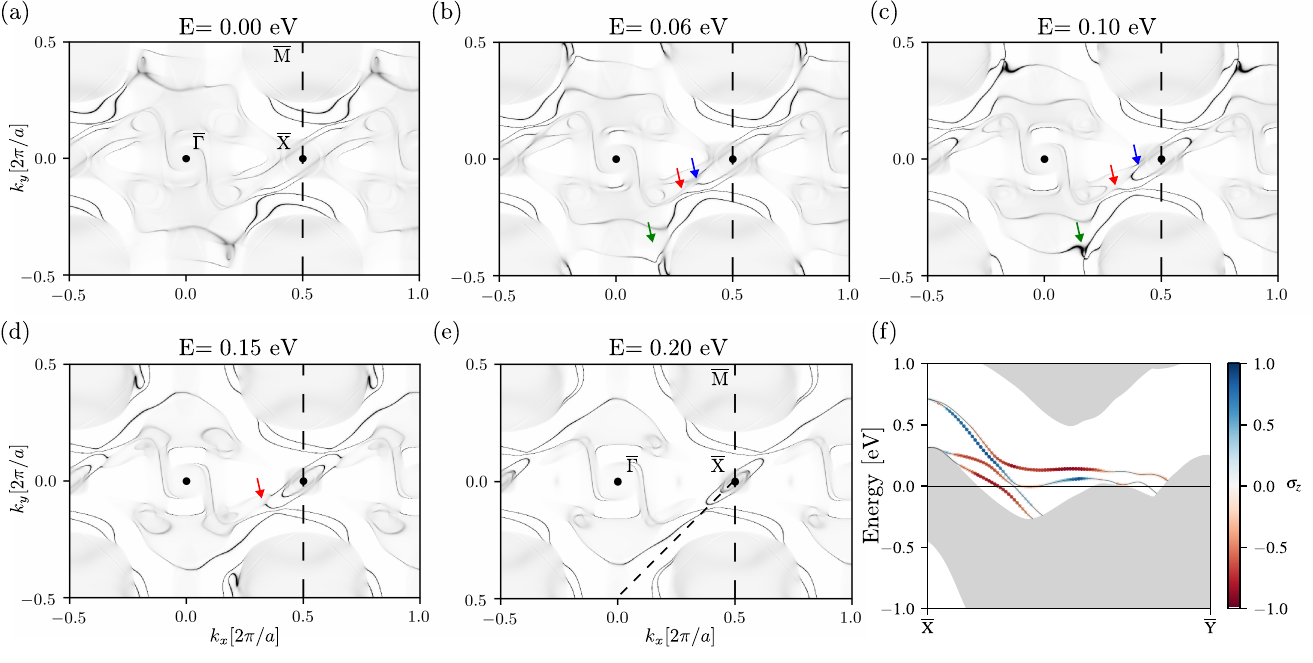}
	\caption{
	Spectral function $A(\mathbf{k}, \omega)$ at fixed energy, evaluated at
	a depth of 10 Bohr radii (5.29~\r{A}) for the (001) surface termination. In panel
	(a), $A(\mathbf{k}, \omega)$ is shown at the Fermi level, while panels
	(b), (c), (d), and (e) correspond to $\omega = 0.006,\ 0.10,\ 0.15$, and
	0.2 eV, respectively. The red, green, and blue arrows highlight the
	change of connectivity between Fermi arcs, as discussed in the main
	text. Panel (f) displays the band structure of a 10-layer slab along the
	dashed path indicated in panel (e). Bulk states are shown in gray, while
	surface states are colored according to the expectation value of
	$\hat{S}_z$. We define a surface state as a band whose atomic weight on
	the first layer exceeds 30\%. Surface states associated with the opposite
	termination are not shown.
	}
	\label{fig:aube_sf}
\end{figure*}

B20 compounds have been shown to host 4-fold degenerate crossings
at the time-reversal-invariant momenta (TRIM) $\Gamma$ and 6-fold ones at R,
with chiral charge of $-4$ and $+4$ respectively~\cite{Yao2020, Bohwmick2025,Mardanya2024,Chang2017}. 
Here, we find them
at $-0.4$~eV (at $\Gamma$) and at $-1.5$~eV (at R), in agreement with
Ref.~\cite{Bohwmick2025}. We further confirm the value of the topological charge
by explicitly integrating the Berry curvature over a small surface centered
around such crossings.
These \( \pm 4 \) Weyl fermions give rise to four surface Fermi arcs connecting \(
\Gamma \) to the projection of R onto the two-dimensional (2D) surface BZ
(denoted by \( \overline{\mbox{M}} \)). The arcs can be visualized by computing the surface spectral function on the (001) termination.
Fig.~\ref{fig:aube_sf} shows the results for different energy cuts starting from
the Fermi level (a) up to 0.2~eV above it (e). 
The best resolution of the arcs is achieved at higher energies,
Fig.~\ref{fig:aube_sf}(e), where fewer bulk bands are present: very long
states originating from \( \overline{\Gamma} \) and following a complex trajectory
culminate in a ``spiraling'' behavior near \( \overline{\mbox{M}} \).
Their connectivity becomes clear if the plot is extended to the neighboring
BZ. Contrary, at the Fermi level, the arcs strongly hybridize with
the surrounding states, and they become hard to track.

To gain deeper insight into the physics of the arcs we examine their
evolution as a function of energy. We observe three
key features, each marked by a different colored arrow in
Fig.~\ref{fig:aube_sf}. Starting form 0.2~eV, the first feature appears around
0.15 eV: it corresponds to the opening of the
circular surface states at the $\overline{\mbox{X}}$ point, the red arrow in panel (d).
Simultaneously, the long arcs around $\overline{\mbox{M}}$ are pushed
downward, signaling the onset of a connectivity exchange between the two. 
This process continues down to approximately 0.10 eV, where the inner circle
around $\overline{\mbox{X}}$ also breaks open (blue arrow).  At 0.06~eV, the
arcs originally stemming from the right side of $\overline{\Gamma}$ are now
attached to the spectral weight around $\overline{\mbox{X}}$, rather than
connecting to $\overline{\mbox{M}}$.
At this point, it is necessary to comment on the additional spectral weight near $\overline{\mbox{X}}$. These supplementary
surface states also originate from the same Fermi arcs connecting
$\overline{\Gamma}$ to
$\overline{\mbox{M}}$: 
their strong dispersion in this compound causes them to cross the
Fermi level multiple times. A similar behavior is theoretically predicted for
GeRh~\cite{Mardanya2024}, but here the effect appears to be significantly
more pronounced.
In panel (f) we also show the spin character of these surface states.
We observe an avoided crossing along the
$\overline{\mbox{X}}$--$\overline{\mbox{Y}}$ direction: the
less dispersive arcs with strong spin-down (red) components are ``separated'' by
the more dispersive spin-up (blue) state. 

Lastly, another connectivity exchange takes place around 0.10~eV, which we
indicate with the green arrow in Fig.~\ref{fig:aube_sf}(b-c). This process is
accompanied by the  formation of a U-shaped structure with significant spectral
weight, clearly visible in panel (c). An analogous behavior has been reported to
occur also in RhSi, and it was found to realize type-II
(surface) van Hove singularities~\cite{Sanchez2023}. It is therefore reasonable to expect the emergence of the same type of singularity in our case. Although we did not explore this aspect further in the present work, it certainly warrants further examinations. Recent experimental~\cite{Kuibarov2024} and
theoretical~\cite{Trama2024,Nomani2023,Bai2025,kuibarov2025} studies have shown that Fermi
arcs can be highly sensitive to superconducting instabilities, leading to
critical temperatures significantly higher than those observed in bulk crystals.
Therefore, BeAu is a promising candidate to further investigate such
phenomenon, given the very large size of the arcs and the potential presence of van Hove singularities.

At this point, we observe that our surface state spectra are broadly consistent with the earlier results of Ref.~\cite{Bohwmick2025}, though some discrepancies are present. We attribute these differences to the choice of structural inputs for the DFT calculations (our use of experimentally determined lattice constants and Wyckoff positions versus the DFT-optimized structure in Ref.~\cite{Bohwmick2025}) as well as the possibility of a different (unspecified) energy shift in the latter work.

In addition to the just discussed $|C|=4$ fermions,  BeAu exhibits another type of multi-fold band
crossing near the Fermi level, Fig.~\ref{fig:e_str}(f). This crossing is located at the TRIM point M,
around 0.32~eV, and is referred to as a \textit{double WP} or
\textit{charge--2 Dirac point} (DP). We will use the latter term throughout.
It comprises two overlapping WPs of the same chirality,
resulting in a total topological charge of two. Its existence stems from the
presence of screw-rotation symmetries and time-reversal symmetry (\( \mathcal{T}
\)). As discussed later in Sec.~\ref{sec:unpaired_crossings}, the combined
action of \( \tilde C_i \) and \( \mathcal{T} \) enforces a two-fold degeneracy along
the boundary of the BZ. Consequently, at the TRIM point M, a four-fold
degeneracy must occur. A numerical analysis of the Wannier bands confirms that all
crossings at the M point are indeed charge-2 DPs. The one closest to the Fermi
level carries a topological charge of \( C = -2 \).  
Remarkably, the charge-2 DP at M does not give rise to additional
surface Fermi arcs (see Sec.~\ref{sec:unpaired_crossings}).

We conclude this section by discussing another degeneracy 3.5 eV below the Fermi level, which deserves mentioning as it constitutes the only known case where a band crossing can carry a topological charge of five. Originally, Bradlyn \textit{et al.}~\cite{Barry2016} argued that the maximum Chern number permitted by crystalline symmetries originates from a 4-fold crossing adiabatically connected to the spin-$3/2$ Hamiltonian, yielding $|C|=4$ at half-filling.
More recently, Alpin \textit{et al.}~\cite{Alpin2023} showed that in space groups 195 and 199, a 4-fold crossing can exist that is \textit{not} adiabatically connected to the spin-$3/2$ model, allowing Chern numbers of $(3, -5, +5, +3)$ for the respective bands. However, the $C = \pm 5$ bands are always paired with  the ones carrying $C = \mp 3$, so the net topological charge at half filling remains $\pm 2$, preserving the effective upper bound of $|C|=4$.
In Fig. 1, panel (d), we color the band with \( C = -5 \) in purple and the band
with \( C = +3 \) in green. A similar 4-fold crossing occurs at even lower
energies.

\subsection{Unpaired Weyl crossings}
\label{sec:unpaired_crossings}

\noindent 
Since periodicity requires that 
no net flux of Berry curvature crosses the surface of the BZ, nodes with
equal and opposite chirality usually come in pairs, resulting in an overall net chirality of
zero. 
To satisfy the condition of zero net Berry curvature flux across the BZ while
permitting an isolated WP, a compensating source or sink of Berry
curvature must exist. In certain chiral crystal structures, rather than a
partner WP, this role is played by \textit{nodal
walls}: symmetry-protected nodal surfaces~\cite{Wu2018} that reside on the
boundary of the BZ.

In such systems, the conventional picture of Weyl semimetals as stacks of 2D
Chern insulators in reciprocal space becomes inadequate. Due to the presence of
nodal surfaces at \textit{all} BZ boundaries, no 2D slice of the
three-dimensional BZ exhibits a full band gap. Consequently, the existence of a partner node is no longer required as the zero net-flux condition is met because of the nodal walls.
Similarly, bulk-boundary correspondence breaks down, and such isolated WPs are not accompanied by topologically protected surface
states~\cite{Yu2019}.

In BeAu, the combination of time-reversal symmetry and skew rotations pins the
nodal degeneracies at the edges of the BZ, hence the name nodal walls. This is
because $(\mathcal{T} \tilde C_{i})^2 = -e^{-i\mathbf k \cdot \mathbf t_i}$,
where $\mathbf t_i$ refers to a translation along $i$.  Thus, Kramers'
degeneracies are guaranteed at all $k_i = \pi$ planes since $(\mathcal T \tilde
C_i)^2 = -1$.
These degeneracies can be seen in the band structure shown in
Fig.~\ref{fig:e_str}(a-b), for example, along the M--X path. We marked them by
red arrows for convenience. Note that as one moves away from the BZ
boundary towards the interior (e.g., along $\Gamma$--M), the bands split again.
It is important to stress that only the bands merging into these nodal walls can
host unpaired WPs. For instance, the fourfold-degenerate fermions at
$\Gamma$ and R originate from bands not joined by nodal walls and thus behave as conventional pairs
of Weyl nodes with associated surface Fermi arcs~\ref{sec:multifold}. 

\begin{figure}[h]
        \centering
        \includegraphics[width=0.5\textwidth]{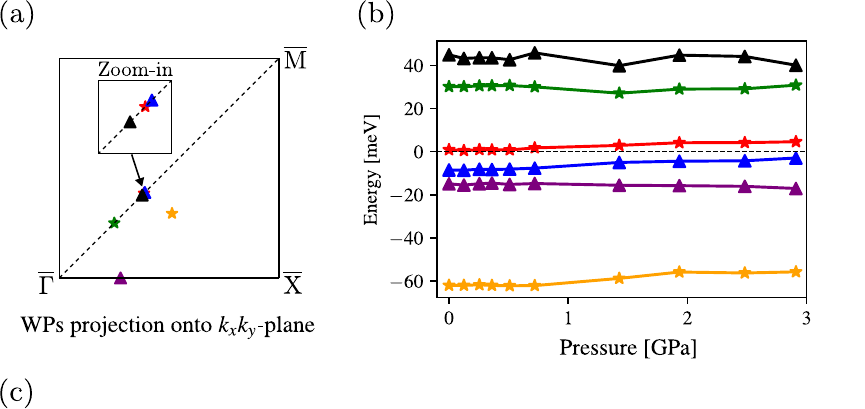}
        \centering
        \begin{tabular}{|c|c|c|c|}      
        \hline  
        ($k_x,k_y,k_z$) & $C$ & Energy [eV] & band number \\
        \hline  
 $0.1928,0.1928,0.1928$ & $\nm 1 $ & $\nm 0.0009 $ & $50\quad (N-2)$ \\ 
 $0.1251,0.1251,0.1251$ & $\nm 1 $ & $\nm 0.0300 $ & $50\quad (N-2)$ \\
 $0.2565,0.1469,0.2319$ & $\nm 1 $ & $   -0.0620 $ & $51\quad (N-1)$ \\ 
 $0.1946,0.1946,0.1946$ & $-1$     & $   -0.0085 $ & $51\quad (N-1)$ \\ 
 $0.1394,0.0000,0.1219$ & $-1$     & $   -0.0148 $ & $51\quad (N-1)$ \\ 
 $0.1887,0.1887,0.1887$ & $-1$     & $\nm 0.0449 $ & $51\quad (N-1)$ \\ 
        \hline  
        \end{tabular}
        \caption{Unpaired WPs at generic $k$-points in the energy window of
                $\sim \pm 50$~meV around the Fermi level. (a) Projection onto
                the $k_z=0$ plane of the WP positions belonging to the
                irreducible BZ. The asterix (triangle) marks correspond to
                positive(negative) chirality. (b) WPs' energy dependece
                on external pressure. (c) Exact 3-dimensional position (units of
                $2 \pi / a$) of the nodes, with corresponding energy and
                chirality.  The ``band number'' refers to our Wannier-projected
                Hamiltonian. We further label the same bands according to the
		order in which they appear at the four-fold fermion at $\Gamma$
		in Fig.~\ref{fig:e_str}(a): $N+2$ corresponds to the highest eigenvalue. 
	}
        \label{fig:ch_4:wps}
\end{figure}

The unpaired crossings can be conventional Weyls or even
multi-fold fermions and can occur both at high-symmetry points---such
as the charge-2 Dirac point at M---as well as at generic
points in reciprocal space. To systematically explore their distribution, we
performed a dense-mesh scan of the full BZ. Since there is no gap in the
energy spectrum up to several eV from the Fermi level, we restricted our
attention to the bands which intersect the Fermi level,
identifying 81 WPs in the irreducible octant of the first BZ.  We report in
Fig.~\ref{fig:ch_4:wps}~(c) the energy and positions of six unpaired WPs found within an
energy window of 50 meV.  Their projection onto the $k_xk_y$-plane is also shown
in panel (a) of the same figure.  

In Fig.~\ref{fig:ch_4:wps} (b), we show the energy dependence of the crossings
under applied pressure, which remains largely unchanged up to 3 GPa. This
behavior aligns with the minimal variation observed in the DOS reported 
in the literature~\cite{arxiv}. To determine the crystal structure at different
pressures, we used the experimental curves for the lattice parameter $a$ as a
function of pressure from Ref.~\cite{arxiv} and 
performed scalar-relativistic symmetry-constrained relaxation of the atomic coordinates.
This step was essential because even tiny changes in the crystal structure have a significant impact on the energy of the nodes. For example, the WP at $-0.062$~eV in our plot lies outside the
energy range of $\pm50$~meV, however, using the value of $a_0 = 4.6699$~\AA\
from Ref.~\cite{Amon2018} yields roughly $-0.040$~eV. 

\section{Fermi Surface}
\label{sec:fs}

\noindent 
In this section we present a detailed analysis of the Fermi surface of BeAu
focusing on the topological charge carried by each part of the
surface. Our motivation is twofold. First, the presence of
unconventional crossings, we find, leads to Fermi sheets carrying unexpectedly
high topological charge. Second, this quantity, also referred as the Chern number
of a sheet, together with the sign of the superconducting gap characterizes the topology of a time-reversal invariant superconductor via Eq.~\eqref{eq:invariant}~\cite{Qi2010,Hosur2014}.

\begin{figure}[ht]
\centering
\includegraphics[width=0.44\textwidth]{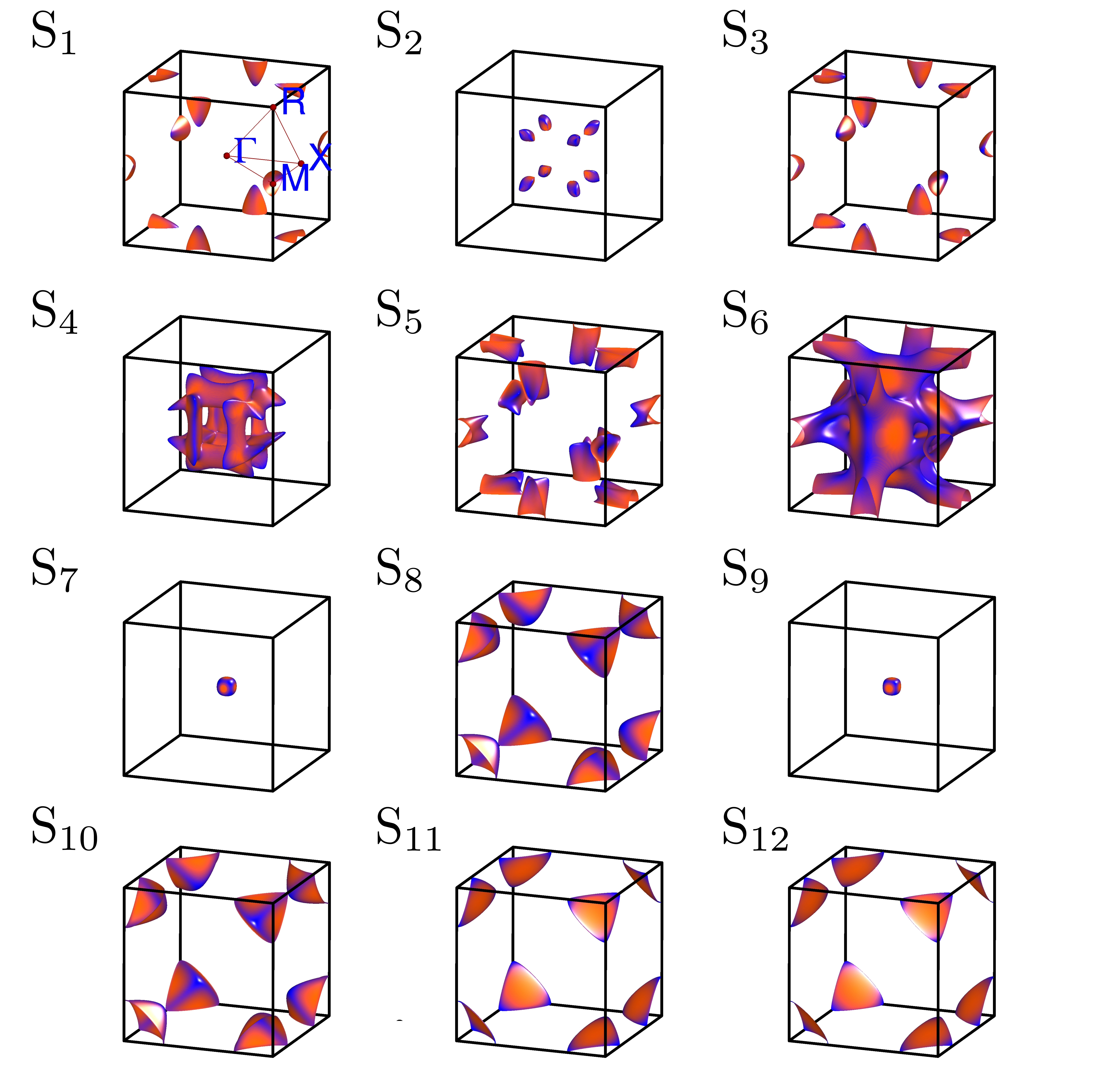}
\caption{All full-relativistic Fermi surfaces of BeAu computed with GGA and SOC with the
de Haas--van Alphen module of FPLO. We used a $24\times24\times24$ $k$-mesh with additional 3
bisections. The color gradient represents the Fermi velocity intensity (units not
shown), ranging from low (blue) to high (orange) values.}
\label{fig:all_fs_sheets}
\end{figure}

First of all, we present the Fermi surface in Fig.~\ref{fig:all_fs_sheets},
which is highly complex, featuring numerous nested as well as intersecting
sheets.
We label each one increasingly from 1 to 12 in the order shown in the figure.
In our notation, S$_1$, S$_3$, and S$_5$ each consist of three pockets, meaning
they technically correspond to three distinct sheets for each plot. We
group them together because they are symmetry-related.
Around \(\Gamma\), there are two small concentric and closed Fermi surfaces (S$_7$
and S$_9$), which
become degenerate in the absence of SOC.  An analogous situation arises near the
R point, where, however, there are two pairs of intersecting surfaces
($\mbox{S}_8 \cup \mbox{S}_{10}$ and $\mbox{S}_{11} \cup \mbox{S}_{12}$). In this second case the spin-orbit splitting is hardly
noticeable.  

To characterize the Chern number of each sheet we perform a numerical
integration to explicitly compute the flux:
\begin{align}
	\begin{split}
	C_i &= \frac 1 {2 \pi} \oint_{\mbox{S}_i} \Omega_i(\mathbf k) \cdot d^2
	\mathbf k \\
	    &= \frac 1 {2 \pi} \oint_{\mbox{S}_i} \Omega_i(\mathbf k) \cdot
	    \mathbf {\hat n}_i(\mathbf k) d^2k, 
	\end{split}
	\label{eq:chern_number}
\end{align}
where S$_i$ denotes the fully connected $i$-th part of the Fermi surface and
$\Omega_i(\mathbf k) = \nabla_{\mathbf k} \times \bra{u_{i,\mathbf k}} i\partial_{\mathbf k}
\ket{u_{i,\mathbf k}}$ is the Berry curvature evaluated at the surface S$_i$.
The vector $\mathbf {\hat n}_i$ specifies the normal direction of the surface
element $d^2\mathbf k$ and it points along the direction of the Fermi velocity. 
We further recall that when multiple surfaces intersect only the flux summed over all
touching regions has physical significance and can be expected to be quantized.

In conventional Weyl semimetals, Eq.~\eqref{eq:chern_number} can be evaluated
implicitly, without explicit integration: a sheet enclosing a
WP possesses a Chern number $|C| > 0$, whereas if no WPs are
enclosed, the sheet has $C = 0$. This procedure, although still theoretically
applicable here, becomes challenging due to the abundance of unpaired WPs and nodal surfaces.

To numerically perform the integration, we first obtain a fine triangulation of
each surface S$_i$. Then the Berry curvature over each plaquette is obtained by
averaging the values of $\Omega_i(\mathbf k)$ at the three vertices. Finally, the
normal vector $\mathbf {\hat n}_i$ is computed from the coordinates of the
plaquettes' vertices with simple geometric rules.
The main practical challenge in evaluating Eq.~\eqref{eq:chern_number} arises for intersecting sheets. In such cases, two kind of errors become relevant when sampling the Fermi surfaces: (i) Wannier interpolation is not exact, and (ii) the Fermi surface is obtained by interpolating nearby points, so the k-points assigned to a given surface $S_i$ do not strictly lie on the iso-energy contour. This introduces a small shift $\delta \mathbf k$, which propagates into the surface integral. These errors are typically negligible since $\Omega_i(\mathbf k)$ varies smoothly, but near band touchings, where $\Omega_i(\mathbf k)$ can change rapidly, even small shifts accumulate. The more degeneracies present, the greater the computational effort required to control them.
To address this issue, we can move slightly away from the Fermi level, where problematic degeneracies are absent or easier to resolve. For instance, this approach was used to determine the Chern number of $S_6$, which was particularly challenging. At just 0.1 eV above the Fermi level, the surface splits into eight pockets centered around X, which we could evaluate separately and then sum their contributions. Finally, since gauge invariance requires $\sum_{i \in \text{FS}} C_i = 0$~\cite{Haldane2004},the chirality can still be inferred from the sum rule in the case one part of the Fermi surface fails to converge.
\begin{figure}[h]
	\centering
	\includegraphics[width=0.48\textwidth]{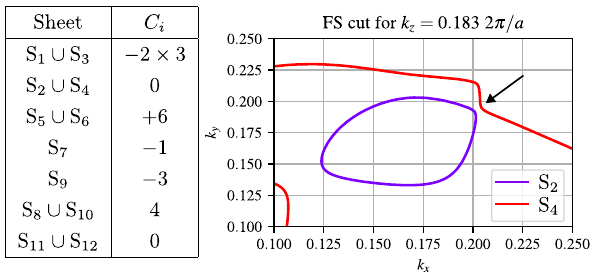}
	\caption{On the left the table summarizes the Fermi surface's Chern
	number as computed numerically from Eq.~\eqref{eq:chern_number}. In the
first line we make explicit that S$_1$ and S$_3$ have multiplicity of 3 because
there are 3 symmetry-related M points. For bookkeeping purposes, we have group them
together in all previous plots. In the second line, we decided to group S$_2$
and S$_4$ together because it was too demanding to computationally resolve the
very small band gap (shown on the right).}
	\label{fig:TABLE}
\end{figure}

The results are summarized with the table in Fig.~\ref{fig:TABLE}.
Interestingly, the flux through the Fermi sheets around $\Gamma$, R and M
matches the value expected from the presence of multifold fermions sitting at the
corresponding high-symmetry points, despite the presence of additional nearby
crossings. 
The two sheets S$_2$ and S$_4$ present  an extremely small gap, shown on the left of Fig.~\ref{fig:TABLE},
which makes direct computation of the flux very sensitive to the mesh. Raising
the energy slightly allows to better resolve the gap and we
conclude (with a reasonable error of $\sim 0.37$~\%) that both of them are
trivial. However, for this conclusion to hold also at the Fermi level, we must know
whether these bands intersect, not only at the Fermi level, but at all energies
between 0 and 0.1~eV otherwise a possible
redistribution of the topological charge is allowed. Given the very small size of the gap, the computational
cost is high and we decided not to pursue this any further. At present, we can say for certain that $C_2 + C_4 = 0$.

\section{Implications for superconductivity}
\label{sec:sc}

\noindent 
Now that all sheets have been characterized, we can comment on the effect of an $s_{\pm}$ pairing state. To investigate the consequences of a sign change in the gap function, Huang et al.~\cite{Peccato2021} assumed a functional form $\Delta(\mathbf{k}) \sim \sum_i \cos(k_i)$. 
{This choice is not unique since any basis of the trivial representation (A) can be chosen—$\cos(k_i)$ being the simplest. While this choice is suitable for their effective model, it need not hold for BeAu.} 

In the absence of further experimental inputs, it is difficult to propose a well-motivated alternative, however, what is important for the appearance of a topological superconducting phase is a sign change between different Fermi surfaces; this could happen in a multi band superconductor if the Fermi sheets involved are, for example, the ones at $\Gamma$ and R. {Under this assumption,} our numerical evaluation of the Fermi surface’s Chern number confirms that a topological superconducting phase with $\nu = 4$ can appear.
We emphasize that numerically evaluating the surface Chern number was
necessary due to the abundance of WPs located at generic
positions, some even very close to $\Gamma$.

Although quite exotic, the presence of M
pockets in theory enables the possibility of a different topological phase with
$\nu = 6$. While such a phase may be unlikely to occur in BeAu due to the very small separation between S$_6$ and S$_4$, similar
electronic structures are common in B20 compounds, making this an intriguing
possibility worth considering.

As a concluding remark, looking at the
atomic-resolved band structure in Fig.~\ref{fig:e_str}, we can also conclude that S$_5$ and S$_6$ {(centered at M)} carry much more Be-character compared to the
neighboring bands at the Fermi level. In Ref.~\cite{arxiv}, a simple model for
the electron-phonon coupling involving only Be was proposed in order to explain the
measured $T_c$. Given the good agreement with the experimental data, 
it is worth keeping in mind
that the observed multi-gap formation could be linked
to a non-uniform distribution of Be-character across the Fermi surface.
{In this case, however, a sign change between $\Gamma$ and R becomes unlikely since the $\Gamma$ and R point have comparable degree of atomic hybridization. On the other hand, the possibility of a sign change between $\Gamma$ and M remains, yielding the previously mentioned $\nu = 6$ topological phase. Such phase was previously commented to be unlikely due to the small separation between the M pocket and other nearby Fermi surfaces but it could be possible. Then, a more involved choice of the basis function for the A-irrep must be invoked to described the $s_{\pm}$ gap function.  Further investigation, i.e. $k$-resolved electron-phonon calculations, are therefore necessary.}

\section{Conclusions}

\noindent 
In this work, we employed DFT combined with Wannierization to characterize the
electronic structure of the chiral Weyl semimetal BeAu. This material belongs to
the B20 family of compounds, which is known for featuring
multi-fold band degeneracies with high Chern numbers.
We first reviewed the band structure of BeAu and identified the presence of
$C=4$ multifold fermions at the $\Gamma$ and R points, consistent with previous
literature.  These band crossings lead to four extended Fermi arcs on the (001)
surface termination, with extremely rich and complex connectivity behavior.

The non-symmorphic skew-rotation symmetries inherent to space group $P2_13$
also permit the presence of unpaired WPs: topological crossings without
counterparts of opposite chirality. In this case, the topological charge is
compensated by nodal surfaces  pinned at the BZ boundaries due to time-reversal
symmetry.
We identified many of these points close to the Fermi level and studied their behavior under hydrostatic pressure up to 3 GPa. We also pointed out that the charge-2 Dirac point at M represents an example of such isolated crossings. 

Motivated by the experimental evidence for multi-gap superconductivity in BeAu and theoretical works on  $s_{\pm}$ pairing in B20 compounds,
we computed the Chern numbers of each Fermi surface sheet. We determined that the sheets surrounding the multifold
fermions at $\Gamma$ and R indeed carry Chern numbers of $\pm 4$, consistent with a $\nu = 4$
topological superconducting phase~\cite{Peccato2021}.
In the course of our analysis, we also discovered a large Fermi surface region
carrying a Chern number of +6, the highest reported for a single Fermi sheet.
This suggests the potential for a $\nu= 6$ topological phase, although its
realization remains unlikely. However, similar Fermi surfaces are shared by many
other B20 compounds {(like the fellow superconductor RhGe)}, making our study of broader
relevance.
Additionally, independent of superconductivity, the existence of such
high Chern number Fermi surfaces may have implications for transport
responses beyond those seen in classical Weyl systems.

Finally, our atomic-projected band structure analysis revealed a significant
difference in the orbital composition of the Fermi pocket around M compared
other states at the Fermi level, with the former showing a much stronger Be
character. Given that superconductivity in BeAu is likely enhanced by Be-based
electron-phonon coupling, this insight could be important for understanding the
origin of its multi-gap superconducting behavior.

\section{Acknowledgements}

\noindent 
We thank Eteri Svanidze for stimulating discussions and Ulrike Nitzsche for technical assistance.
This work was supported by the Deutsche Forschungsgemeinschaft (DFG, German Research Foundation) under Germany’s Excellence Strategy through the W\"{u}rzburg-Dresden Cluster of Excellence on Complexity and Topology in Quantum Matter – \textit{ct.qmat} (EXC 2147, project-ids 390858490, 392019).
M.G.V. thanks support to PID2022-142008NB-I00
project funded by MICIU/AEI/10.13039/501100011033
and FEDER, UE, and Canada Excellence Research Chairs
Program for Topological Quantum Matter.
D.~V. received funding from the National Research, Development and Innovation Office of Hungary under OTKA grant no. FK 146499, and
the János Bolyai Research Scholarship of the Hungarian Academy of Sciences.

\bibliography{mybib}

\end{document}